УДК 331.5+004.05

**Табаков Валерій Зіновійович**
*к.т.н., доц., доцент Інституту підготовки кадрів ДСЗУ*

# СИСТЕМНА ПОМИЛКА ПРОЕКТУ ЄТОНН ТА ЄІАС

*Виявлено системну помилку реалізації проекту ЄТОНН та ЄІАС.NET, яка полягає у використанні розробниками ЄІАС.NET операційної системи Windows у складі програмного забезпечення ЄІАС.NET, яке відповідно до вимог технічного завдання на проектування ЄІАС.NET має забезпечити інформаційний обмін в режимі реального часу. Зроблено висновок про неможливість виконання вимоги технічного завдання на проектування ЄІАС.NET щодо забезпечення роботи ЄТОНН/ЄТНаСП за підтримки ЄІАС/ЄІАС.NET в режимі реального час за умови використання операційної системи Windows у складі програмного забезпечення ЄІАС.NET.*

ЄТОНН, ЄТНаСП, ЄІАС, ЄІАС.NET, система реального часу, інформаційна система, державна служба зайнятості

Державні служби зайнятості світу, в тому числі України, виконують п'ять груп основних функцій:

- посередництво у працевлаштуванні (трудове посередництво);
- розвиток інформаційних систем ринку праці;
- управління програмами регулювання ринку праці;
- управління системою допомоги по безробіттю;
- управління регуляторними діями.

1999 рік був відзнаменований переходом Державної служби зайнятості України (ДСЗУ) на роботу за Єдиною технологією обслуговування незайнятого населення (ЄТОНН), 2003 – впровадженням Єдиної інформаційно-аналітичної системи (ЄІАС), а 2011 – переходом на нову версію ЄІАС – ЄІАС.NET.

Аналіз етапності становлення служби зайнятості, стану ринку праці надає підстав для висновку про закономірність переходу до роботи ДСЗУ на новій програмній платформі ЄІАС – платформі .NET. Однією з основних вимог до нової платформи, як платформи, на якій реалізується система програмної підтримки технології ЄТОНН, є її спроможність забезпечити роботу ЄІАС в режимі реального часу, і є необхідною вимогою до технології ЄІАС і ЄТОНН.

Починаючи з 2012 року, в ДСЗУ впроваджується удосконалена ЄТОНН зі зміненою назвою: Єдина технологія надання соціальних послуг (ЄТНаСП), що означає розповсюдження технології ЄТОНН та ЄІАС не лише на обслуговування безробітних та роботодавців, а також на обслуговування інвалідів, пенсіонерів та інших верств населення, яким мають надаватися соціальні послуги.

**Постановка проблеми.** Щодо характеристик технології ЄТНаСП та системи ЄІАС.Net як технології і системи, які мають забезпечити надання соціальних послуг в режимі реального часу, то характеристика щодо забезпечення роботи в режимі реального часу залишається однією з основних характеристик ЄТНаСП та ЄІАС.Net і умовою, яка має входити до переліку вимог технічного завдання на проектуваня програмного забезпечення ЄІАС.Net. Реалізація ЄТОНН (ЄТНаСП) за підтримки ЄІАС (ЄІАС.NET) здійснювалась з основною метою: забезпечити зникнення черг безробітних до центрів зайнятості. Проте аналіз фактичної функціональності нових технології та системи не надає права оминути суттєві зауваження (претензії) до роботи ЄІАС.Net з боку фахівців центрів зайнятості з приводу працездатності ЄІАС.Net, а відтак і ЄТОНН (ЄТНаСП), які стосуються забезпечення визначеності граничного часу формування відкликів системи на інформаційний запит, та підставність віднесення ЄІАС.Net до системи, яка має працювати, в режимі реального часу.

**Аналіз найновіших публікацій.** В основу ЄТНаСП та ЄІАС покладена єдина технологія надання соціальних послуг центрами зайнятості України, розроблена авторським колективом Інституту підготовки кадрів державної служби зайнятості України у складі Ю.М.Маршавіна, Л.М.Фокас, Л.Є.Лямiної, М.М.Руженського та інших [1].

У кожному центрі зайнятості, незалежно від площі, планування приміщень та інших умов, організуються десять функціональних секторів. У кожному секторі можуть виконувати свої функції спеціалісти як одного підрозділу, так і декількох підрозділів центру зайнятості відповідно до своїх службових обов'язків, структури конкретного центру зайнятості, що має забезпечити якісне надання послуг і повне дотримання даної технології. Сектори можуть бути розміщені в окремій кімнаті, декількох кімнатах, частині кімнати, великій залі, холі, вестибюлі, коридорі. Просторове розміщення секторів здійснюється відповідно до технологічного ланцюжка – послідовності роботи з клієнтом (рис. 1). Формування функціональних секторів необхідне для оптимізації роботи спеціалістів ЦЗ та їх взаємодії і тому не потребує спеціального позначення для привернення уваги відвідувачів.

Перелік секторів у центрах зайнятості:

I – довідковий;
II – самостійного пошуку вакансій;
III – профінформаційний;
IV – запобігання соціальним ризикам;
V – підбору роботи;
VI – активної підтримки безробітних;
VII – інформування роботодавців;
VIII – самостійного пошуку роботодавцями претендентів на роботу;
IX – співпраці з роботодавцями;
X – навчання та психологічного розвантаження персоналу;
XI – адміністративно-господарський.

Відразу після входу до центру зайнятості розташовується довідковий сектор. Спеціаліст, який виконує довідково-диспетчерські функції, як правило, перебуває поблизу вхідних дверей у приміщенні центру зайнятості і першим зустрічає клієнта. Наступними мають бути сектори самостійного пошуку вакансій, профінформаційний та запобігання соціальним ризикам, а вже потім – сектор підбору роботи та сектор активної підтримки безробітних. Сектори інформування роботодавців, самостійного пошуку роботодавцями претендентів на роботу, співпраці з роботодавцями, адміністративно-господарський, навчання та психологічного розвантаження персоналу можуть розташовуватися навіть в іншій частині приміщення центру зайнятості і мати окремий вхід, оскільки вони пов'язані з виконанням особливих, специфічних функцій.

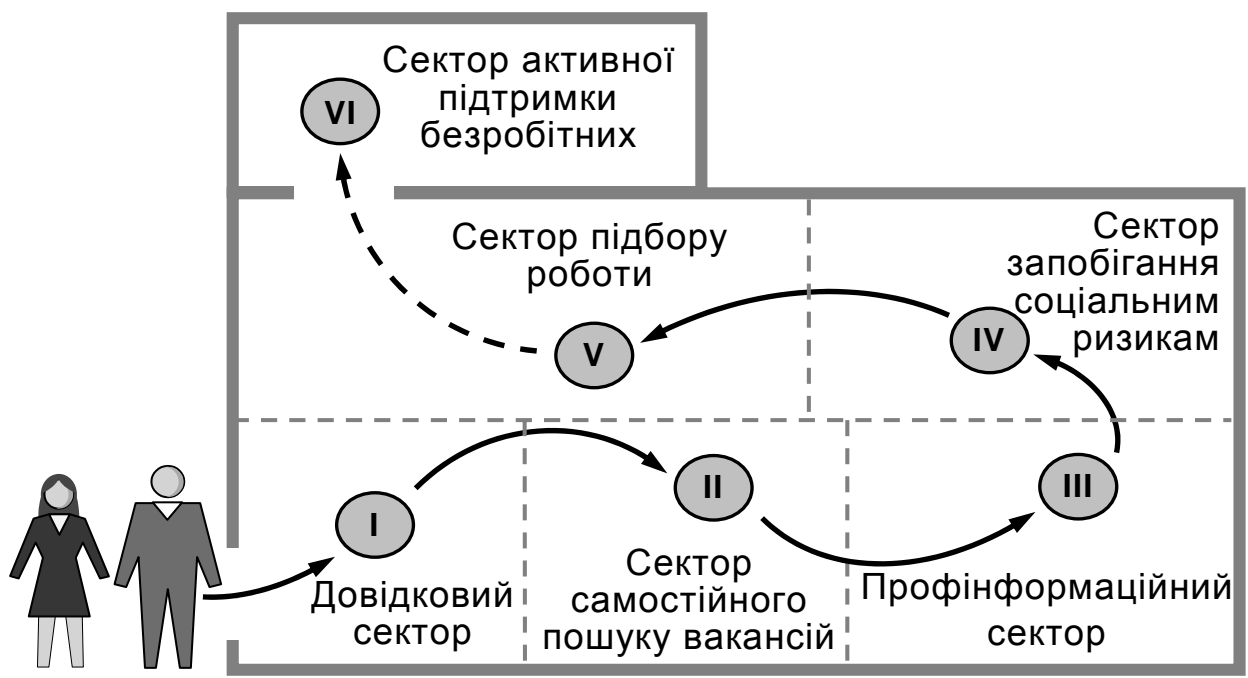

**Рис. 1**. Структурно-логічна схема обслуговування шукачів роботи в центрах зайнятості

**Постановка проблеми та мета дослідження.** Просторово-часовий розподіл технологічного ланцюжка ЄТНаСП має забезпечити

• безперервність обслуговування клієнтів спеціалістами центрів зайнятості;

• чітке дотримання розрахованого похвилинно графіку обслуговування клієнтів;

• відповідний темп підтримки цього графіку засобами ЄІАС.Net в режимі реального часу.

Таким чином ми робимо висновок про обґрунтованість вимоги до програмного забезпечення ЄІАС. Net, щодо забезпечення роботи ЄІАС. Net в режимі реального часу.

Про характеристику впровадженої в діяльність ДСЗУ ЄІАС.Net щодо роботи в режимі реального часу неодноразово наголошувалось у публікаціях в засобах масової інформації, інтернет-порталі ДСЗУ та інших джерелах. Проте, фактично ЄІАС.NET не може бути кваліфікована як система реального часу, що є наслідком системної помилки її проектувальників. Виявлення наявності системної помилки реалізації проекту ЄТНаСП&ЄІАС.NET та її причин складає проблему та мету нашого дослідження.

**Матеріал та методика.** Комп'ютерна система, що включає технічні пристрої, операційну систему та прикладне програмне забезпечення називається системою реального часу, якщо для такої системи можна гарантувати деяке наперед відоме максимально можливе значення часу реагування системи на надходження даних (інформаційний запит).

Характеристичною особливістю систем реального часу є гарантований час виконання будь-якої операції, завдяки чому можливе виконання програм в заданому темпі.

Необхідно розуміти, що система реального часу не обов'язково забезпечує малий час виконання тих або інших операцій. Справа не в конкретному розмірі проміжку часу, а в тому, що максимально можливий розмір цього проміжку гарантований.

Система реального часу може бути в середньому навіть повільнішою, ніж могла би бути за інших рівних умов звичайна система загального призначення, проте система реального часу гарантує виконання операцій в строго визначений час.

За спостереженнями Кудряшова І.Г., наведеними у [2], у сучасній практиці реалізації програмних систем виявлено наступне: "На самом деле, сейчас системы общего назначения зачастую используются в тех областях, где следует использовать только системы реального времени. Скажем, во многих технологических установках программное обеспечение работает под управлением операционной системы Microsoft Windows NT, которая системой реального времени не является. Для того чтобы снизить вероятность недопустимо больших задержек в работе разработчики таких систем вынуждены проектировать аппаратную часть с учетом возможности задержек и устанавливать значительно (иногда в несколько раз) более быстродействующие компьютеры. Но даже такие действия, строго говоря, не гарантируют от сбоев в работе систем. Эта практика, безусловно, порочна, но достаточно распространена."

ЄІАС.Net принципово не здатна забезпечити роботу в режимі реального часу, оскільки містить в контурі управління обчислювальними та інформаційними процесами складові, що є частинами операційної системи Windows, які в процесі інсталяції налаштовуються автоматично, що виключає можливість забезпечити виконання умови, що є обов'язковою для системи реального часу, а саме умови визначеності максимального часу реагування системи на запит.

Нездатність ЄІАС.Net забезпечити роботу в режимі реального часу є причиною того, що в процесі експлуатації ЄІАС.Net виникають черги не обслугованих запитів, що в свою чергу унеможливлює дотримання похвилинно розрахованого графіку обслуговування клієнтів за ЄТНаСП та спричиняє утворення черг клієнтів в центрах зайнятості.

Ця особливість спроектованої ЄІАС.Net стала більш очевидною тоді, коли база даних ЄІАС.Net була повністю розміщена на сервері Державного центу зайнятості України.

**Висновки та перспективи подальших досліджень.** Таким чином нами виявлено системну помилку реалізації проекту ЄТНаСП&ЄІАС.NET, яка полягає у використанні розробниками ЄІАС.NET операційної системи Windows у складі програмного забезпеченя ЄІАС.NET, яке відповідно до вимог технічного завдання на проектування ЄІАС.NET має забезпечити інформаційний обмін в режимі реального часу. Використання операційної системи Windows у складі програмного забезпеченя ЄІАС.NET, за будь-яких умов, навіть за умов використання супершвидкіних, а відповідно і супердорогих, апаратних пристроїв, унеможливлює забезпечення вимоги технічного завдання на проектування ЄІАС.NET щодо забезпечення роботи ЄТОНН (ЄТНаСП) за підтримки ЄІАС (ЄІАС.NET) в режимі реального часу.

Вважаю, що відповідальність за помилку в проекті ЄТОНН&ЄІАС.NET повністю лежить на авторах ЄТОНН (ЄТНаСП), а також на фахівцях аутсорсингової компанії – розробника ЄІАС.Net ЗАТ «Софтлайн» Ользі Фокас та інших.

Чи є ознаки корупції у діях посадових осіб, внаслідок чого у проекті ЄТОНН (ЄТНаСП) &ЄІАС (ЄІАС.NET) зроблено системну помилку?

Чи зробить керівництво Державного центру зайнятості висновки за результатами виявленої причини невідповідності ЄТНаСП&ЄІАС.Net вимогам проекту?

Чи буде виправлено системну помилку реалізації проекту ЄТНаСП&ЄІАС.Net?

Чи отримаємо ми відповіді на ці запитання?

Література:

Табаков Валерий Зиновьевич, **Системная ошибка проекта ЕТОНН и ЕИАС.**

*Выявлена системная ошибка реализации проекта ЕТОНН та ЕИАС.NET, которая состоит в использовании разработчиками ЕИАС.NET операционной системы Windows в составе программного обеспечения ЕИАС.NET, которое в соответствии с требованиями технического задания на проектирование ЕИАС.NET должно обеспечивать информационный обмен в режиме реального времени. Сделан вывод о невозможности выполнения требования технического задания на проектирование ЕИАС.NET про обеспечение работы ЕТОНН/ЕТНаСП при поддержке ЕИАС/ЕИАС.NET в режиме реального времени при условии использования операционной системы Windows в составе программного обеспечения ЕИАС.NET.*

ЕТОНН, ЕТНаСП, ЕИАС, ЕИАС.NET, система реального времени, информационная система, государственная служба занятости

Tabakov Valery Zinovievich. **System error of project *ETONN and EIAS.***

*Detected a system error of the project ETONN&EIAS.NET, which consists in the use of the developers EIAS.NET the Windows operating system in the software EIAS.NET, which in accordance with the requirements of the specification for the project EIAS.NET should provide information exchange in real time . It is concluded that it is impossible to meet the requirement specification for the project EIAS.NET about providing work ETONN/ETNaSP with the support of EIAS / EIAS.NET in real time, providing the use of the Windows operating system in the software EIAS.NET.*

ETONN, ETNaSP, EIAS, EIAS.NET, real-time system, information system, the public employment service